\documentclass[aps,prl,twocolumn,amsmath,showpacs]{revtex4}
\usepackage{bm}
\usepackage{graphicx}
\begin{document}
\setlength{\arraycolsep}{2pt}
\title{Uncertainty inequalities as entanglement criteria for negative partial-transpose states}
\author{Hyunchul Nha$^{1,*}$ and M. Suhail Zubairy$^{1,2}$} 
\affiliation{$^1$Department of Physics, Texas A \& M University at Qatar, Doha, Qatar\\
$^2$Department of Physics and Institute of Quantum Studies, Texas A\& M University, College Station, TX 77843, USA}
\date{\today}
\begin{abstract}
In this Letter, we show that the fulfillment of uncertainty relations is a {\it sufficient} criterion for a quantum-mechanically permissible state.
We specifically construct two pseudo-spin observables for an arbitrary non-positive Hermitian matrix whose uncertainty relation is violated.
This method enables us to systematically derive separability conditions for all negative partial-transpose states in experimentally accessible forms.
In particular, generalized entanglement criteria are derived from the Schr{\"o}dinger-Robertson inequalities for bipartite continuous-variable states.
\end{abstract}
\pacs{03.65. Ta, 03.65.Ud, 03.67.Mn, 42.50.Dv}
\maketitle

\narrowtext
Quantum mechanics sets a bound on the product of uncertainties of two non-commuting observables at the fundamental level. This uncertainty principle must be fulfilled as a necessary condition for a quantum physical state \cite{Manko0}. In the present Letter, we want to take a deeper view of the role of uncertainty relations by asking: Can the satisfaction
of uncertainty relations be regarded as a {\it sufficient} condition for a legitimate quantum state?

Our question can be possibly rephrased in many different forms the answers to which may each provide us with valuable insight to quantum physics. One of them is: Can there be any non-positive Hermitian operator that satisfies {\it all} uncertainty relations?
In fact, this fundamental issue was addressed by a number of people for decades, e.g., in \cite{Narcowich}, particularly in the phase-space framework of quantum mechanics \cite{Moyal}. Notably, it has been claimed that there {\it is} a certain non-positive Hermitian operator that fulfills the uncertainty relations. No one, however, ever presented a conclusive argument, as all considered only a restricted class of
uncertainty relations involving the canonical variables in the lowest order. In this Letter, we demonstrate that the satisfaction of uncertainty relations for {\it all} pairs of noncommuting observables is indeed {\it sufficient} to represent a quantum physical state.
In particular, given an {\it arbitrary} non-positive
Hermitian matrix, we explicitly construct two pseudo-spin observables whose uncertainty relation is violated.

Besides its fundamental importance, our explicit construction has an immediate application as entanglement criteria for the whole class of negative partial-transpose (NPT) states in arbitrary dimensions. When a given state $\rho$ is separable, it is written as a convex sum of product states, $\rho=\Sigma_ip_i\rho_1^{(i)}\otimes\rho_2^{(i)}\cdots\otimes\rho_N^{(i)}$,
where the state $\rho_j^{(i)}$ refers to the subsystem $j$. Under partial transposition (PT) for a set of subsystems,
a separable state still remains positive, therefore it describes a certain physical state \cite{Peres}. A number of entanglement
criteria have been derived based on PT, and remarkably, all the known criteria for continuous variables (CVs)
belong to this category \cite{Shchukin,Agarwal,nha1,nha2,Serafini,Hillery}. In particular, the uncertainty
relations under PT were employed as the necessary condition for separability \cite{Agarwal,nha1,nha2,Serafini,Guhne}.
Up to now, however, it was not clear to what extent the uncertainty-relation-based approach can detect entangled states.
Moreover, given a general {\it mixed} entangled state, it is nontrivial to identify the inequality that can be violated
by the state.

In this respect, our construction remarkably shows that
the uncertainty-relation-based approach is in fact {\it sufficient} to detect bipartite entanglement for the whole class of NPT states.
More importantly, it enables us to {\it systematically} derive entanglement conditions for a given NPT state in experimentally accessible forms.
We also show that a weaker form of our nonlinear inequality is equivalent
to a well-known entanglement witness formalism \cite{Horodecki}, providing the witness operator with a physical interpretation as such.
We specifically address the generalization of separability conditions based on the Schr{\"o}dinger-Robertson inequalities for CVs \cite{SR1},
which recovers as special cases the previously known criteria,
and illustrate their utility in detecting entangled states generated via beam splitter.

Let us start by introducing uncertainty relations. Given two non-commuting observables, $\{A,B\}$,
the widely-known Heisenberg uncertainty relation (HUR) reads $\Delta A \Delta B \ge\frac{1}{2}|\langle[A,B]\rangle|$.
Less known is the more generalized Schr{\"o}dinger-Robertson (SR) inequality \cite{SR1},
\begin{eqnarray}
\langle(\Delta A)^2\rangle\langle(\Delta B)^2\rangle\ge\frac{1}{4}|\langle[A,B]\rangle|^2+\frac{1}{4}
\langle\Delta A\Delta B\rangle_S^2,
\label{eqn:SR}
\end{eqnarray}
where the covariance $\langle\Delta A\Delta B\rangle_S$ is defined in a symmetric form,
$\langle\Delta A\Delta B\rangle_S\equiv\langle\Delta A\Delta B+\Delta B\Delta A\rangle.$ Clearly, the SR inequality generally
provides a stronger bound on the product of uncertainties than the HUR \cite{nha2}.

First, we consider the simplest case of 2-dim systems, which provides us with a valuable insight to the current issue.
A general $2\times2$ Hermitian matrix $\rho$ is given in a form
\begin{eqnarray}
\rho=\begin{pmatrix}
&a&c\\&c^*&b
\end{pmatrix},
\end{eqnarray}
($a$, $b$: real, $c\equiv c_r+ic_i$: complex).
For this matrix to represent a physical state, two conditions must be met: (i) ${\rm Tr} \{\rho\}=a+b=1$ and
(ii) ${\rm Det}\left[\rho\right]=ab-|c|^2\ge0$. Throughout the present paper, we assume that the trace condition is met,
${\rm Tr} \{\rho\}=1$, which can be relaxed later. Then, the only remaining condition is (ii), which turns out to be just a
SR-inequality in Eq.~(\ref{eqn:SR}): Take the angular momentum operators $S_i=\frac{\hbar}{2}\sigma_i$, where $\sigma_i$ is
the Pauli spin operator $(i=x,y,z)$. Then, 
one obtains
$(\Delta S_x)^2(\Delta S_y)^2=\frac{\hbar^4}{16}(1-4c_r^2)(1-4c_i^2)$, $\langle[S_x,S_y]\rangle=i\frac{\hbar^2}{2}(a-b)$,
and $\langle\Delta S_x\Delta S_y\rangle_S=2\hbar^2c_rc_i$. On inserting these results to Eq.~(\ref{eqn:SR}),
one immediately finds $ab-|c|^2\ge0$, the condition (ii). Here, the use of the SR inequality is important as one would instead have
$ab-|c|^2+4c_r^2c_i^2\ge0$ through the HUR \cite{note,nha3}.

Therefore, we have the proposition that the physical realizability for 2-dim systems is equivalent to the satisfaction of the
{\it single} SR inequality between $S_x$ and $S_y$.
At this point, it is worthwhile to observe that
the two Hermitian operators $S_x$ and $S_y$ are represented using the basis states, $|0\rangle$ and $|1\rangle$,
as $S_x=\frac{\hbar}{2}\left(|0\rangle\langle1|+|1\rangle\langle0|\right)$ and $S_y=\frac{\hbar}{2i}
\left(|0\rangle\langle1|-|1\rangle\langle0|\right)$, which will be used below to construct two pseudo-spin observables
to our end.

Now, let us turn our attention to a general Hermitian matrix $\rho$ of arbitrary dimension $N$ with
${\rm Tr} \{\rho\}=1$. In general, $\rho$ has the real eigenvalues $\lambda_i$ and the corresponding eigenstates
$|\lambda_i\rangle$ $(i=1,\cdots,N)$, i.e. $\rho|\lambda_i\rangle=\lambda_i|\lambda_i\rangle$, with the orthonormality condition
$\langle\lambda_i|\lambda_j\rangle=\delta_{ij}$. Due to the trace condition, ${\rm Tr} \{\rho\}=\Sigma_i\lambda_i=1$, there always
exists at least one positive eigenvalue for $\rho$. 

Let us define two pseudo-spin observables $H_1$ and $H_2$ in the Hilbert space spanned by two eigenstates
$|\lambda_1\rangle$ and $|\lambda_2\rangle$ as
\begin{eqnarray}
H_1=\alpha_1|\lambda_1\rangle\langle\lambda_2|+\alpha_1^*|\lambda_2\rangle\langle\lambda_1|\nonumber\\
H_2=\alpha_2|\lambda_1\rangle\langle\lambda_2|+\alpha_2^*|\lambda_2\rangle\langle\lambda_1|,
\label{eqn:observ}
\end{eqnarray}
where $\alpha_1$ and $\alpha_2$ are complex constants. 
Denoting $x\equiv{\rm Re}(\alpha_1\alpha_2^*)$ and $y\equiv{\rm Im}(\alpha_1\alpha_2^*)$, the commutator $[H_1,H_2]=2iy
\left(|\lambda_1\rangle\langle\lambda_1|-|\lambda_2\rangle\langle\lambda_2|\right)$ and the anticommutator
$\{H_1,H_2\}=2x\left(|\lambda_1\rangle\langle\lambda_1|+|\lambda_2\rangle\langle\lambda_2|\right)$
follow together with $H_i^2=|\alpha_i|^2\left(|\lambda_1\rangle\langle\lambda_1|+|\lambda_2\rangle\langle\lambda_2|\right)$
($i=1,2$). As $\rho$ is diagonal in the eigenstate basis, $\rho={\rm diag}\{\lambda_1,\lambda_2,\cdots,\lambda_N\}$,
it is straightforward to show
\begin{eqnarray}
\langle(\Delta H_i)^2\rangle&=&\langle H_i^2\rangle=|\alpha_i|^2(\lambda_1+\lambda_2)\nonumber\\
\langle[H_1,H_2]\rangle&=&2iy(\lambda_1-\lambda_2)\nonumber\\
\langle\Delta H_1\Delta H_2\rangle_S&=&2x(\lambda_1+\lambda_2).
\label{eqn:exp}
\end{eqnarray}
The SR inequality is now reduced to 
\begin{eqnarray}
4y^2\lambda_1\lambda_2\ge0.
\end{eqnarray}
The above inequality is clearly violated if $\lambda_1\lambda_2<0$, that is, the case that one eigenvalue is positive and
the other negative. Therefore, we deduce the fact that the satisfaction of {\it all} SR inequalities is sufficient and necessary to endorse a Hermitian matrix of
unit trace as a legitimate quantum state.

Note that one could relax the unit-trace condition to any positive trace
if the normalization of the matrix were allowed.
Furthermore, in Eq.~(\ref{eqn:exp}), we observe another interesting point for a non-positive Hermitian matrix.
If two or more eigenvalues are negative, there exists an observable, $H_i$ $(i=1,2)$, whose variance becomes negative. This is of course a clear
signature of being unphysical, as all physical observables must have nonnegative variances. Therefore, {\it  for a general trace-class Hermitian operator, the satisfaction of all uncertainty relations together with the positivity of variances is sufficient and necessary
as a legitimate quantum state.}

Aside from its fundamental importance, our explicit construction of the two observables in Eq.~(\ref{eqn:observ})
can have some practical applications.
One of them is the derivation of entanglement condition on demand for an NPT state in arbitrary dimensions. Given a certain $N$-partite
state $\rho$, one may wish to determine whether the system possesses bipartite entanglement between two parties, one party $S_1$
composed of the subsystems $1,\cdots,j$ and the other $S_2$ of the subsystems $j+1,\cdots,N$. Suppose that the state is biseparable as
$\rho=\Sigma_ip_i\rho_{S_1}^{i}\otimes\rho_{S_2}^{i}$. Then by taking transposition only on $S_2$,
the density operator transforms as $\rho^{\rm PT}=\Sigma_ip_i\rho_{S_1}^{i}\otimes(\rho_{S_2}^{i})^{\rm T}$,
which is still positive
definite. In other words, a separable state still remains physical under PT. All the uncertainty relations must therefore be
fulfilled as
\begin{eqnarray}
&&\langle(\Delta H_1)^2\rangle_{\rm PT}\langle(\Delta H_2)^2\rangle_{\rm PT}\nonumber\\
&&\ge\frac{1}{4}|\langle[H_1,H_2]\rangle_{\rm PT}|^2+\frac{1}{4}\langle\Delta H_1\Delta H_2\rangle_{S,{\rm PT}}^2,
\label{eqn:SRPT}
\end{eqnarray}
where the subscript $\rm PT$ denotes the quantum average over $\rho^{\rm PT}$ as $\langle{\hat O}\rangle_{\rm PT}\equiv{\rm Tr}
\{\hat O\rho^{\rm PT}\}$ (${\hat O}$: arbitrary operator).
The inequality~(\ref{eqn:SRPT}) must be satisfied by a separable state for any arbitrary operators $H_1$ and $H_2$, and it thus becomes a separability condition in general \cite{nha2}.

Note that PT preserves the trace and the hermicity of the density operator.
Then, for a general entangled state that has some negative eigenvalues under PT,
{\it there always exists at least one uncertainty relation that is violated by the NPT state.} This means that
uncertainty-relation-based approach to detection of bipartite entanglement is sufficient for the whole
class of NPT states.
Furthermore, our construction in Eq.~(\ref{eqn:observ}) enables us to derive an uncertainty inequality
as separability condition for a given NPT state as follows.

Given an NPT state $\rho$, one first obtains the eigenvalues and the corresponding eigenstates for the PT density operator.
Then take any two eigenstates for $\lambda_1>0$ and $\lambda_2<0$, and construct the two observables as $H_1=\frac{1}{2}
\left(|\lambda_1\rangle\langle\lambda_2|+|\lambda_2\rangle\langle\lambda_1|\right) $ and $H_2=\frac{1}{2i}\left(|\lambda_1\rangle
\langle\lambda_2|-|\lambda_2\rangle\langle\lambda_1|\right)$.
Then, the SR inequality in Eq.~(\ref{eqn:SRPT}) is violated by the given state.
Eq.~(\ref{eqn:SRPT}) can be further expressed
in terms of quantum averages of a normal density operator using the general relation ${\rm Tr}\{{\hat O}\rho^{\rm PT}\}
={\rm Tr}\{{\hat O}^{\rm PT}\rho\}$, or
\begin{eqnarray}
{\rm Tr}\{|ij\rangle\langle i'j'|\rho^{\rm PT}\}={\rm Tr}\{|ij'\rangle\langle i'j|\rho\},
\end{eqnarray}
where $|i\rangle$ and $|i'\rangle$ are the basis states for party $S_1$, and $|j\rangle$ and $|j'\rangle$ the ones for party $S_2$.
The inequality~(\ref{eqn:SRPT}) can then be expressed in terms of the observables measured with respect to the given state $\rho$,
instead of $\rho^{\rm PT}$.


As an illustration, let us consider a class of tripartite mixed state $\rho_{\rm GHZ}=p|GHZ\rangle\langle GHZ|+\frac{1-p}{8}I$,
where $|GHZ\rangle=\frac{1}{\sqrt{2}}\left(|000\rangle+|111\rangle\right)$. With respect to bipartition \{AB,C\}, the density
operator under PT has two different eigenvalues, $\lambda_+=\frac{1+3p}{8}$ (degenerate) and $\lambda_-=\frac{1-5p}{8}$, thus
becomes an NPT state for $p>\frac{1}{5}$. Taking two eigenstates for $\rho_{\rm GHZ}^{\rm PT}$, $|\lambda_{\pm}\rangle
=\frac{1}{\sqrt 2}\left(|001\rangle\pm|110\rangle\right)$, one obtains two observables $H_1=\frac{1}{2}\left(|\lambda_+\rangle\langle\lambda_-|+
|\lambda_-\rangle\langle\lambda_+|\right)=\frac{1}{2}\left(|001\rangle\langle001|-|110\rangle\langle110|\right)$ and
$H_2=\frac{1}{2i}\left(|\lambda_+\rangle\langle\lambda_-|-|\lambda_-\rangle\langle\lambda_+|\right)=\frac{1}{2i}
\left(|110\rangle\langle001|-|001\rangle\langle110|\right)$. Then, using the method outlined above, the separability condition
is obtained as
\begin{eqnarray}
(4A_z-B_z^2)(4A_z-C_{xy}^2)\ge 16D_{xy}^2+B_z^2C_{xy}^2,
\end{eqnarray}
where $A_z\equiv\langle I+\sigma_{1z}\sigma_{2z}-\sigma_{1z}\sigma_{3z}-\sigma_{2z}\sigma_{3z}\rangle$, $B_z\equiv
\langle \sigma_{1z}+\sigma_{2z}-\sigma_{3z}-\sigma_{1z}\sigma_{2z}\sigma_{3z}\rangle$, $C_{xy}\equiv\langle
\sigma_{1x}\sigma_{2x}\sigma_{3y}+\sigma_{1x}\sigma_{2y}\sigma_{3x}+\sigma_{1y}\sigma_{2x}\sigma_{3x}-\sigma_{1y}\sigma_{2y}
\sigma_{3y}\rangle$, and $D_{xy}\equiv\langle \sigma_{1x}\sigma_{2x}\sigma_{3x}-\sigma_{1x}\sigma_{2y}\sigma_{3y}-\sigma_{1y}
\sigma_{2x}\sigma_{3y}-\sigma_{1y}\sigma_{2y}\sigma_{3x}\rangle$.
The inequalities for the other two bipartitions are obtained simply by taking permutations, and all those separability conditions
are violated by the state $\rho_{\rm GHZ}$ for $p>\frac{1}{5}$, thereby characterizing to some extent its tripartite entanglement.

The above method holds valid for any NPT states in arbitrary dimensions.
One can also derive a weaker form of separability condition as
\begin{eqnarray}
\langle H_1^2\rangle_{\rm PT}\langle H_2^2\rangle_{\rm PT}\ge\frac{1}{4}|\langle[H_1,H_2]\rangle_{\rm PT}|^2.
\label{eqn:WHURPT}
\end{eqnarray}
The inequality in~(\ref{eqn:WHURPT}) is weaker than the one in~(\ref{eqn:SRPT}),
as $\langle(\Delta H_i)^2\rangle_{\rm PT}\le \langle H_i^2\rangle_{\rm PT}$ $(i=1,2$):
If the inequality~(\ref{eqn:WHURPT}) is violated, so is the inequality~(\ref{eqn:SRPT}), but the converse is not always true \cite{note1}.
It is now straightforward to show that the inequality in~(\ref{eqn:WHURPT}) is reduced to
${\rm Tr }\{W_1\rho\}{\rm Tr}\{W_2\rho\}\ge0$, where $W_i=|\lambda_i\rangle\langle\lambda_i|^{\rm PT}$.
Since $\lambda_1$ is taken as a positive eigenvalue for $\rho^{\rm PT}$, it is further reduced to ${\rm Tr}\{W_2\rho\}\ge0$,
which is none other than the formalism of the entanglement witness \cite{Horodecki}.
In other words, the class of entanglement witness operator based on the PT of the entangled state $|\lambda_2\rangle$
can be given a physical interpretation as a weaker form of uncertainty inequality.

Note that it is always more advantageous to use the {\it nonlinear} form of the inequality~(\ref{eqn:SRPT}) than the linear entanglement witness \cite{Horodecki},
as the former can be more robust against experimental noises in practice.
Moreover, one can also show that any pair of two orthogonal vectors with the condition $\langle\lambda_1|\rho|\lambda_1\rangle>0$ and $\langle\lambda_2|\rho|\lambda_2\rangle<0$, not necessarily eigenvectors as used so far, can be employed in constructing two pseudo spin observables in Eq.~(\ref{eqn:observ}).
As a consequence, a richer class of separability inequalities may be derived for a given NPT state, which will be addressed elsewhere.

In principle, the above method can also be applied to CVs which can include both finite- and infinite-dimensional systems. For finite-dimensional CV states, e.g., the single-photon entangled state $\frac{1}{\sqrt{2}}\left(|0\rangle|1\rangle+|1\rangle|0\rangle\right)$ \cite{Hillery}, our method can generally work as a practical tool to derive the separability condition. For infinite-dimensional CV states, however, except for symmetric states, e.g. EPR state,
the computation of the eigenstates may be less tractable than finite-dimensional systems.
From another perspective, nevertheless,
one can establish an alternative approach still relying on the uncertainty principle.
For instance, using a general form of two-mode state $|\lambda\rangle=\Sigma C'_{mn}a_1^{\dag m}a_2^{\dag n}|0,0\rangle$ and
$|0\rangle\langle0|_i=:\hspace{-0.14cm}e^{-a_i^\dag a_i}\hspace{-0.14cm}:$ ($i=1,2$, :: denotes normal-ordering),
two general observables $H_1$ and $H_2$ can be expressed
in terms of the dyadic $|\lambda_1\rangle\langle\lambda_2|=\Sigma\frac{(-1)^{k+k'}}{k!k'!}
C_{mn}D_{pq}a_1^{\dag (m+k)}a_1^{n+k}
a_2^{\dag (q+k')}a_2^{p+k'}$ and its conjugate $|\lambda_2\rangle\langle\lambda_1|$.
Using the relation
$\langle a_1^{\dag m}a_1^na_2^{\dag p}a_2^q\rangle_{\rho^{PT}}=\langle a_1^{\dag m}a_1^na_2^{\dag q}a_2^p\rangle_{\rho}$
\cite{Shchukin,nha1},
one can derive the separability conditions via the uncertainty relation of $H_1$ and $H_2$.
Remarkably, the satisfaction of them for arbitrary $C_{mn}$ and $D_{pq}$
is {\it sufficient and necessary} for the separability of two-mode NPT states.

Instead of pursuing a full generalization, left for future work, we demonstrate
here that even a little generalization can work out a wide class of important inequalities for CVs.
We illustrate the utility of those inequalities by detecting two-mode entanglement generated via a beam-splitter.
It is known that a single-mode nonclassical state is sufficient to generate an entangled state
via 50:50 beam splitter with the other input in vacuum state \cite{Asboth}, however, the entanglement detection at the output
is another nontrivial issue to resolve. In fact, it was also conjectured that the whole class of entangled states
via beam-splitter is NPT, which is, though, not yet rigorously proved \cite{Simon}.
In this respect, the uncertainty-relation-based approach is very relevant
to such class of entangled states.

Let us first define $X_i^{(m)}\equiv a_i^{\dag m}+a_i^m$ and $Y_i^{(m)}\equiv -i(a_i^{\dag m}-a_i^m)$ for two modes $i=1,2$,
and take two Hermitian operators, $H_1=X_1^{(m)}+X_2^{(n)}$ and $H_2=Y_1^{(m)}+Y_2^{(n)}$.
The separability condition then follows from the SR inequality, as
\begin{eqnarray}
\Delta^2H_1\Delta^2{\tilde H}_2
\ge\left<C_1^{(m)}+C_2^{(n)}\right>^2
+\langle\Delta H_1\Delta {\tilde H}_2\rangle_S^2,
\label{eqn:cv1}
\end{eqnarray}
where ${\tilde H}_2\equiv Y_1^{(m)}-Y_2^{(n)}$, and $C_i^{(m)}\equiv[a_i^{m},a_i^{\dag m}]$ ($i=1,2$).
For $m=n=1$, a weaker form of the above inequality, i.e., the HUR version ignoring the last term in (\ref{eqn:cv1}), is reduced to the one derived by Mancini {\it et al.} \cite{Mancini} which is known to be stronger than
the ones by Duan {\it et al.} \cite{Duan}.
For another class of inequalities, take $H_1=a_1^{\dag m}a_2^{\dag n}+a_1^{m}a_2^{n}$ and
$H_2=-i(a_1^{\dag m}a_2^{\dag n}-a_1^{m}a_2^{n})$. Then, one obtains
\begin{eqnarray}
&&\left(\Delta^2X_{mn}+\langle C_1^{(m)}C_2^{(n)}\rangle\right)\left(\Delta^2Y_{mn}+\langle C_1^{(m)}C_2^{(n)}\rangle\right)\nonumber\\
&&\ge\left<\left[a_1^{m}a_2^{n},a_1^{\dag m}a_2^{\dag n}\right]
\right>^2+\langle\Delta X_{mn}\Delta Y_{mn}\rangle_S^2,
\label{eqn:cv2}
\end{eqnarray}
where $X_{mn}\equiv a_1^{\dag m}a_2^n+a_1^{m}a_2^{\dag n}$, $Y_{mn}\equiv -i(a_1^{\dag m}a_2^n-a_1^{m}a_2^{\dag n})$.
For $m=n=1$, the HUR version of (\ref{eqn:cv2})
is reduced to the one in \cite{Agarwal,nha1}, and the SR version to the one in \cite{nha2}.
Furthermore, for general $m$ and $n$, the sum form of HUR version, which is generally weaker than the product form \cite{nha1}, is reduced to the class of inequalities in \cite{Hillery}.

The above inequalities in (\ref{eqn:cv1}) and (\ref{eqn:cv2}) can successfully detect
general two-mode entanglement out of a beam-splitter. Using the inequality (\ref{eqn:cv1}), one can detect entanglement generated by the whole class
of nonclassical states with arbitrary-order amplitude squeezing \cite{Shchukin2},
$\langle:(\Delta (a^me^{-i\phi}+a^{\dag m}e^{i\phi}))^2:\rangle <0$.
$m=1$ case refers to the
normal quadrature squeezing and $m=2$ to the amplitude-squared squeezing \cite{Hillery2}. On the other hand, the inequality
(\ref{eqn:cv2}) detects entanglement by the class of arbitrary-order nonclassical photon statistics,
$\langle:(\Delta a^{\dag m}a^m)^2:\rangle <0$ ($m=1$: sub-Poissonian).
An efficient experimental scheme was proposed to measure general correlation functions of arbitrary orders
in \cite{Shchukin3}, which may be suitable for the test of above inequalities.

Finally, let us briefly discuss how the entanglement detection based on the uncertainty relations can be connected to the formalism
by Shchukin and Vogel \cite{Shchukin}. For a positive Hermitian operator $\rho$,
the condition ${\rm Tr}\{\hat {F}^{\dag}\hat {F}\rho\}\ge0$ must be met
for an arbitrary operator $\hat {F}$.
Shchukin and Vogel have derived a hierachy of sufficient and necessary conditions for the positivity under PT
by considering a general form of ${\hat F}$.
A practical difficulty, though, would be to single out an adequate condition violated by a given state among all of them.
On the other hand, we showed that only a special class
of the operator ${\hat F}=c_1\Delta H_1+c_2\Delta H_2$ is sufficient to detect non-positivity,
for which ${\rm Tr}\{\hat {F}^{\dag}\hat {F}\rho\}\ge0$ is reduced to the SR inequality~(\ref{eqn:SR}) by requiring positiveness for any $c_i$'s $(i=1,2)$.
Remarkably, our "state-specific" method, which is powerful particularly for finite-dimensional systems, directly relates a single separability condition to any given state in a more physically intuitive term, uncertainty principle.


HN acknowledges Q. Sun, J. Bae, C.~Noh, and H. J. Carmichael for helpful discussions. 
This work is supported by an NPRP grant from the Qatar National Research Fund.
*email: hyunchul.nha@qatar.tamu.edu


\begin{references}
\bibitem{note} However, if one applies an arbitrary unitary operation to the two observables $S_x$ and $S_y$ and then requires that the HUR must be fulfilled in each case, the same condition (ii) emerges. In this sense, the set of all HURs is unitarily
equivalent to a single SR inequality.
\bibitem{note1} The condition $\langle(\Delta H_i)^2\rangle_{\rm PT}\ge0$ ($i=1,2$) is assumed in this argument.
If, on the other hand, $\langle(\Delta H_i)^2\rangle_{\rm PT}<0$ holds, it is a signature of quantum entanglement by itself.
\end{references}
\end{document}